\documentclass[10pt, a4paper]{article}

\makeatletter \@addtoreset{equation}{section} \makeatother

\usepackage[cp1251]{inputenc}
\usepackage{amsmath}
\usepackage{feynmp}
\usepackage{wrapfig}
\usepackage{amssymb}
\usepackage{graphicx}
\usepackage{appendix}
\usepackage{slashed}
\usepackage{indentfirst}

\begin{document}

\begin{center}
\Large\textbf{Cherenkov radiation in moving medium}
\end{center}

\medskip

\begin{center}
Mihail Alfimov$^{1,2}$

${}^{1}$\emph{P.N. Lebedev Physical Institute, 117924 Leninsky pr.
53, Moscow, Russia}

${}^{2}$\emph{Department of General and Applied Physics, MIPT, 141700
Institutsky per. 9, Dolgoprudny, Russia}

\end{center}

\begin{abstract}
Cherenkov radiation in uniformly moving homogenous isotropic medium
without dispersion is studied. Formula for the spectrum of Cherenkov
radiation of fermion was derived for the case when the speed of the
medium is less than the speed of light in this medium at rest. The
properties of Cherenkov spectrum are investigated.
\end{abstract}

\section*{Introduction}

Cherenkov radiation is an outstanding phenomenon, which has
applications in a range of various domains. Classical description of
this phenomenon was obtained a long time ago. Also there exist a
number of quantum considerations of this effect. Spectrum of
Cherenkov radiation in the medium at rest was calculated in a number
of papers (see, for example \cite{Manoukian}, \cite{Ryazanov} and
\cite{Schwinger}). There are not so many articles and reviews
regarding Cherenkov radiation in moving medium (see, for example
\cite{Bolotovsky}). However, in \cite{Bolotovsky} only classical
radiation without quantum corrections was considered.

The aim of this article is to obtain the formula of Cherenkov
radiation spectrum in moving medium by systematic calculation in the
framework of quantum in-medium QED. Also we are going to analyze how
the shape of this spectrum and angle distribution of this radiation
depend on the parameters of the considered system.

Calculation of the spectrum of Cherenkov radiation in moving medium
is of some interest because of its relation to phenomena in
nucleus-nucleus collisions. Significant experimental data concerning
an existence of double-humped structure of away-side azimuthal
correlations got in the experiments STAR \cite{STAR1, STAR2} and
PHENIX \cite{PHENIX1, PHENIX2, PHENIX3} at RHIC aroused a new wave of
interest to a phenomenon of Cherenkov radiation of gluons first
considered in \cite{Dremin1,Dremin2,Apanasenko}. In the relation with
this phenomenon it is interesting to analyze Cherenkov radiation in
moving medium because the medium that is formed in nucleus-nucleus
collisions is expanding. It is important that in this situation the
radiating particle and the medium move in the same direction.

\section{Effective notion of medium in QED}

Let us consider the effect of medium for processes in QED. We will
use the simplest model with constant dielectric permittivity and
neglect temporal and spatial dispersion. Then the dielectric
permittivity tensor \cite{Ter-Mikaelyan} in the reference frame, in
which the medium is at rest takes the form

$$\varepsilon^{\rho\sigma}=\begin{pmatrix} \varepsilon & 0& 0& 0\\0& -1& 0& 0\\0& 0& -1&
0\\0& 0& 0& -1\end{pmatrix}.$$

In arbitrary reference frame the dielectric permittivity tensor can
be written as follows

\begin{equation}
\varepsilon_{\mu\nu}=g_{\mu\nu}+(n^2
-1)\widetilde{v}_{\mu}\widetilde{v}_{\nu}, \label{epsilon}
\end{equation}
where $n=\sqrt{\varepsilon}$ is a refraction index, and
$\widetilde{v}^{\mu}=\left( \frac{1}{\sqrt{1-v^2}},
\frac{\overrightarrow{v}}{\sqrt{1-v^2}} \right) $ is a 4-velocity of
the medium and $\overrightarrow{v}$ is a spatial velocity.

The aim of the present article is a calculation of the spectrum of
Cherenkov radiation in moving medium. The calculation will be made in
the Lorentz gauge $\varepsilon_{\mu\nu}\partial^{\mu}A^{\nu}=0$. The
first thing to find out is the photon dispersion law in moving
medium. Equation of motion of the field $A^{\mu}$ in the Lorentz
gauge takes the form
\begin{equation}
\varepsilon_{\mu\rho} \varepsilon_{\nu\sigma}
\partial^{\mu}
\partial^{\rho} A^{\sigma}(x)=0.
\end{equation}
In momentum representation one has, correspondingly
\begin{equation}
\varepsilon_{\mu\rho} \varepsilon_{\nu\sigma} q^{\mu} q^{\rho}
A^{\sigma}(q)=0,
\end{equation}
so that the covariant dispersion law of the photon in medium reads
\begin{equation}
\varepsilon_{\mu\rho} q^{\mu} q^{\rho}=0.
\end{equation}
Substituting the definition of the dielectric permittivity tensor
from (\ref{epsilon}) we have
\begin{gather}
q^2+(n^2-1)(q\widetilde{v})^2=0, \\
\left(1+\frac{n^2-1}{1-v^2}
\right)(q^0)^2-2q^0\frac{n^2-1}{1-v^2}|\textbf{q}|v\cos\theta-|\textbf{q}|^2\left(
1-\frac{n^2-1}{1-v^2}v^2 \cos^2 \theta \right)=0, \label{q_eq}
\end{gather}
where $\theta$ is the angle between the photon momentum $\textbf{q}$
and the velocity of the medium $\textbf{v}$. The quadratic equation
(\ref{q_eq}) has two solutions
$$q^0=\frac{\frac{n^2-1}{1-v^2}v\cos\theta \pm \left( 1+\frac{n^2-1}{1-v^2}(1-v^2\cos^2 \theta) \right)^{\frac{1}{2}} }{1+\frac{n^2-1}{1-v^2}}|\textbf{q}|.$$
The solution with the sign ``$-$`` in the limit $v=0$ transforms to
$q^0=-\frac{|\textbf{q}|}{n}$, the solution with the sign ``$+$`` in
the limit $v=0$ transforms to $q^0=\frac{|\textbf{q}|}{n}$. That is
why we choose the solution with the sign ``$+$`` as physically
sensible. In our case phase velocity $V_{ph}(\textbf{q})$ determined
by the dispersion relation $q^{0}=V_{ph}|\textbf{q}|$ depends on the
direction of the photon propagation
\begin{equation}
V_{ph}=\frac{\frac{n^2-1}{1-v^2}v\cos\theta + \left(
1+\frac{n^2-1}{1-v^2}(1-v^2\cos^2 \theta) \right)^{\frac{1}{2}}
}{1+\frac{n^2-1}{1-v^2}}. \label{Dispersion_law}
\end{equation}
Such a photon dispersion law in medium was derived in
\cite{Bolotovsky}.

\begin{figure}
  \begin{center}
  \includegraphics[width=0.85\linewidth]{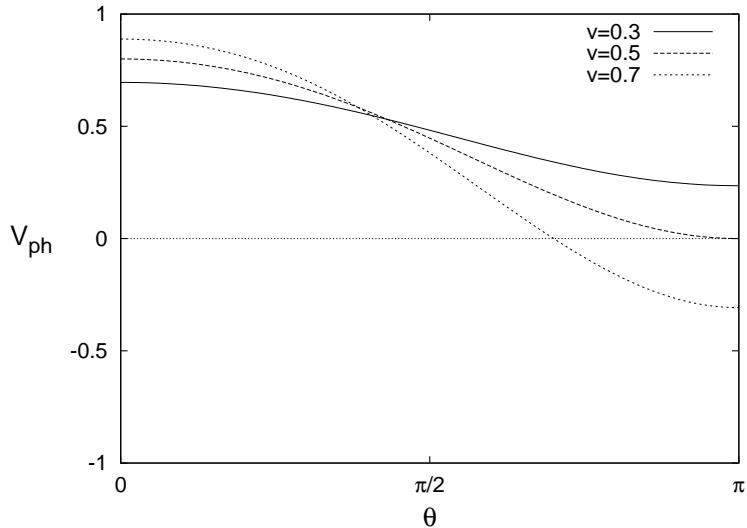}\\
  \end{center}
  \caption{Phase velocity of the photon in medium for $n=2.0$. Solid line $v=0.3$, dashed line $v=0.5$, dotted line $v=0.7$.}\label{ris:disp}
\end{figure}

It is easy to verify that if $|v|<\frac{1}{n}$, the phase velocity is
positive for all $\theta$, but if $|v| \geqslant \frac{1}{n}$, there
exists the region of $\theta$, where $V_{ph}<0$. This statement is
illustrated by the figure \ref{ris:disp}. Below we examine the case,
when the velocity of the medium is smaller, than the phase speed of
light in this medium at rest. Conjointly, one can easily prove that
$V_{ph}<1$ for all $v$ satisfying $|v| <1$.

In what follows we shall need an expression for the photon propagator
$D_{\mu\nu}(q)$ in the Lorentz gauge. According to Ter-Mikaelyan
\cite{Ter-Mikaelyan} it is represented by the formula
\begin{equation}\label{prop_phot}
    D_{\mu\nu}(q)=\frac{-i(g_{\mu\nu}+(n^{-2}-1)\widetilde{v}_{\mu}\widetilde{v}_{\nu})}{q^2+(n^2-1)(q\widetilde{v})^2+i\varepsilon}.
\end{equation}

\section{Spectrum of Cherenkov radiation}

The spectrum of Cherenkov radiation reads (see, for example,
\cite{Schwinger})
\begin{equation}
P(\omega)=\frac{\omega}{E} e^2 \int\frac{d^4 q}{(2\pi)^4}
\delta\left(\omega-V_{ph}(\textbf{q})|\textbf{q}| \right) \mbox{Im}
[i\mathcal{M}(p \rightarrow p)], \label{Cherenkov_spectrum}
\end{equation}
where $\omega$ is the photon energy, $E$ is the energy of the
radiating particle, $p$ -- its 4-momentum, $\mathcal{M}(p \rightarrow
p)$ is the matrix element, the $\delta$-function originates from a
unit frequency projection operator and $V_{ph}(\textbf{q})$ is phase
speed of a photon with momentum $|\textbf{q}|$. The $\delta$-function
in (\ref{Cherenkov_spectrum}) ensures that the radiated photon is on
the mass shell. The spectrum $P(\omega)$ corresponds to probability
distribution of energy radiated in the interval of photon energies
$[\omega, \omega+d\omega]$ per unit time.

Leading order contribution to the imaginary part of the matrix
element $\mathcal{M}(p \rightarrow p)$ is given by the cut of
self-energy diagram.

\setlength{\unitlength}{1mm}

\begin{center}
\begin{fmffile}{diagram}
\begin{fmfgraph*}(50,30)
\fmfleft{i1} \fmfright{o1} \fmf{fermion, label=$p$}{i1,v1}
\fmf{fermion, tension=1/2}{v1,v2} \fmf{fermion, label=$p$}{v2,o1}
\fmf{photon, left, tension=0}{v1,v2} \fmftop{t1} \fmfbottom{b1}
\fmf{dashes, width=.5thin, tension=.5}{t1,v3} \fmf{phantom}{v3,b1}
\fmfiv{label=$q$}{(.4w,.87h)} \fmfiv{label=$p-q$}{(.48w,.48h)}
\end{fmfgraph*}
\end{fmffile}
\end{center}

It is important, that if we considered such a diagram in vacuum,
putting initial and final fermion on the mass-shell, this matrix
element would not give any contribution to the imaginary part. The
decay of fermion into fermion and photon is possible only in the
medium.

The expression for the matrix element with photon and electron
propagators
\begin{equation}
i\mathcal{M}(p \rightarrow p)=\int\frac{d^4 q}{(2\pi)^4}
\overline{u}(p)(-ie\gamma^{\mu})S(p-q)(-ie\gamma^{\nu})u(p)D_{\mu\nu}(q).
\label{matrix_element}
\end{equation}

The expression for the spectrum of Cherenkov radiation in medium at
rest was calculated, for example, in articles \cite{Manoukian} and
\cite{Ryazanov}. Cherenkov radiation spectrum was also derived in
approximation of soft photons in article \cite{Schwinger}. It reads
\begin{equation}
P(\omega | v=0, \beta)=\alpha\omega\beta \left\lbrace 1-\frac{1}{n^2
\beta^2}\left( 1+\frac{\omega}{2E}(n^2-1) \right)^2
+\frac{\omega^2}{2E^2}\frac{n^2-1}{\beta^2} \right\rbrace,
\end{equation}
where $\beta$ is the speed of the fermion. For the angle of Cherenkov
radiation of the photon with energy $\omega$ one has
\begin{equation}
\cos\theta_e=\frac{1}{n\beta}\left( 1+\frac{n^2-1}{2}\frac{\omega}{E}
\right).
\end{equation}
The condition $|\cos\theta_e| \leq 1$ limits the energy of Cherenkov
photons. The following investigation of the properties of Cherenkov
radiation in moving medium also contain the conclusions concerning
the case when the medium is at rest.

Now we return to the formula (\ref{Cherenkov_spectrum}) for Cherenkov
spectrum. Then we substitute the propagator of the photon and the
propagator of the fermion
\begin{equation*}
S(p-q)=\frac{i(\slashed p -\slashed q +m)}{(p-q)^2-m^2+i\varepsilon},
\end{equation*}
in (\ref{matrix_element}), where $\widetilde{v}_{\mu}$ -- 4-velocity
of the medium. The propagator of fermion does not contain any
information about the medium, because only photon field is affected
by the medium. Formula (\ref{Cherenkov_spectrum}) transforms to
\begin{multline}
P(\omega)=\frac{\omega}{E} e^2 \int\frac{d^4 q}{(2\pi)^4}
\delta\left(\omega-V_{ph}|\textbf{q}|\right)\times \\
\times \mbox{Im}\frac{i\overline{u}(p)\gamma^{\mu}(\slashed p
-\slashed
q+m)\gamma^{\nu}u(p)(g_{\mu\nu}+(n^{-2}-1)\widetilde{v}_{\mu}\widetilde{v}_{\nu})}{(q^2+(n^2-1)(q\widetilde{v})^2+i\varepsilon)((p-q)^2-m^2+i\varepsilon)}.
\label{Cherenkov_moving_spectrum_1}
\end{multline}

There we must say that in this article the case, when the velocity of
the medium and the momentum of the particle are co-directed, is
considered for simplicity.

Handling Dirac algebra and denoting the numerator we have
$$
N(p,q,\widetilde{v})=-4E^2 \beta^2 \left[1-\frac{1}{\beta^2}+\left(
1-\frac{1}{n^2}\right)\frac{(1-\beta v)^2}{\beta^2 (1-v^2)}-\right.
$$
$$
-\frac{q^0}{2\beta^2 E}\left\lbrace 1+\frac{1}{n^2}+2\left(
1-\frac{1}{n^2}\right)\frac{1-\beta v}{1-v^2} \right\rbrace+
$$
\begin{equation}
\left. +\frac{|\textbf{q}|}{2\beta E}\left\lbrace
1+\frac{1}{n^2}+2\left( 1-\frac{1}{n^2}\right)\frac{v(1-\beta
v)}{\beta(1-v^2)} \right\rbrace \cos\theta \right],
\end{equation}
where $\theta$ is the angle between the velocity of the medium and
the momentum of the photon $|\textbf{q}|$.

After integrating (\ref{Cherenkov_moving_spectrum_1}) over
$|\textbf{q}|$ we obtain
\begin{gather*}
P(\omega)=\frac{\omega^3}{E\left( 1+\frac{n^2-1}{1-v^2} \right)} e^2 \frac{1}{(2\pi)^3} \int dq^0 \int^{1}_{-1} d(\cos\theta) \frac{1}{V_{ph}^3} N(p,q,\widetilde{v})|_{|\textbf{q}|=\omega/V_{ph}} \times \\
\times \mbox{Im}
\frac{i}{(q^0-q^0_{1+})(q^0-q^0_{1-})(q^0-q^0_{2+})(q^0-q^0_{2-})}.
\end{gather*}
There are four poles in this expression: two in the upper half plane
and two in the lower one (see
(\ref{pole_1}),(\ref{pole_2-}),(\ref{pole_2+})).

To perform the integration over $q^{0}$ we close the integration
contour in the lower half plane and find the residues
\begin{multline} \label{radiation}
P(\omega)=\frac{\omega^3}{E\left( 1+\frac{n^2-1}{1-v^2} \right)} e^2 \frac{1}{(2\pi)^3} \int^{1}_{-1} d(\cos\theta) \frac{1}{V_{ph}^3} \times \\
\times 2\pi \mbox{Im} \left[
\frac{\overline{u}Nu|_{q^0=q^0_{1+}}}{D_{1\varepsilon}}+\frac{\overline{u}Nu|_{q^0=q^0_{2+}}}{D_{2\varepsilon}}
\right],
\end{multline}
\begin{gather}
D_{1\varepsilon}=(q^0_{1+}-q^0_{1-})(q^0_{1+}-q^0_{2+})(q^0_{1+}-q^0_{2-}),
\\
D_{2\varepsilon}=(q^0_{2+}-q^0_{1+})(q^0_{2+}-q^0_{1-})(q^0_{2+}-q^0_{2-}).
\end{gather}

It was proved in Appendix B that $D_{1\varepsilon}$ does not
contribute to the integral over $\cos\theta$. This means that only
one pole $q_0=q_{2+}$ contributes to the final expression. Let's
denote the Cherenkov angle $\theta_{e}$ and introduce convenient
notations
\begin{equation}
x=\frac{\omega}{E}, y=\frac{n^2-1}{1-v^2}
\end{equation}

Then
\begin{equation}
\mbox{Im}\frac{1}{D_{2\varepsilon}}=-\frac{\pi}{2\omega E^2
(1-B(\theta_{e}))\beta\frac{x}{V_{ph}^3}\frac{V_{ph}+\frac{xv}{\beta}\sqrt{y(1-V_{ph}^2)}}{V_{ph}+\sqrt{y(1-V_{ph}^2)}}}\delta(\cos\theta-\cos\theta_{e})
\label{D_2}
\end{equation}

The formula (\ref{radiation}) can thus be rewritten as
\begin{equation}
P(\omega)=\frac{\omega^3}{E\left( 1+\frac{n^2-1}{1-v^2} \right)} e^2
\frac{1}{(2\pi)^2} \int^{1}_{-1} d(\cos\theta) \frac{1}{V_{ph}^3} Im
\left[ \frac{\overline{u}Nu|_{q^0=\omega}}{D_{2\varepsilon}} \right].
\label{radiation_2}
\end{equation}

We substitute the obtained expression (\ref{D_2}) in the expression
(\ref{radiation_2}) for Cherenkov radiation spectrum and integrate by
$\cos\theta$. Expression for Cherenkov angle $\cos\theta_e$
(\ref{Cherenkov_angle}) was obtained in Appendix C. After lengthy
calculations we get the final answer for the spectrum
\begin{multline} \label{Spectrum_of_radiation}
P(\omega | v>0, \beta)=\frac{\alpha\omega\beta}{1+\sqrt{2\left( 1+\frac{n^2-1}{1-v^2}\frac{\omega}{E}\frac{v}{\beta}\left( 1-\frac{v}{\beta} \right) -\sqrt{1+2\frac{n^2-1}{1-v^2}\frac{\omega}{E}\frac{v}{\beta}\left( 1-\frac{v}{\beta} \right)} \right)}}\times \\
\times \left[ 1-\frac{1}{\beta^2}+\left( 1-\frac{1}{n^2} \right)\frac{(1-\beta v)^2}{\beta^2 (1-v^2)}- \right. \\ -\frac{\omega}{2\beta^2 E}\left\lbrace 1+\frac{1}{n^2}+2\left( 1-\frac{1}{n^2} \right)\frac{1-\beta v}{1-v^2} \right\rbrace + \\
+\frac{1}{2}\left\lbrace 1+\frac{1}{n^2}+2\left( 1-\frac{1}{n^2} \right)\frac{v(1-\beta v)}{\beta(1-v^2)} \right\rbrace \frac{1-v^2}{v^2 (n^2-1)} \times \\
\times \left. \left(
1+\frac{n^2-1}{1-v^2}\frac{\omega}{E}\frac{v}{\beta}\left(
1-\frac{v}{\beta} \right)
-\sqrt{1+2\frac{n^2-1}{1-v^2}\frac{\omega}{E}\frac{v}{\beta}\left(
1-\frac{v}{\beta} \right)} \right) \right].
\end{multline}
For the Cherenkov angle we get (see Appendix C)
\begin{equation} \label{Cherenkov_angle_1}
\cos\theta_e=\frac{1+\frac{\beta}{xyv}\left(
1-\sqrt{1+2xy\frac{v}{\beta}\left( 1-\frac{v}{\beta} \right)}
\right)}{\sqrt{v^2+\frac{2\beta^2}{x^2}\left( \frac{xv}{\beta}\left(
1-\frac{v}{\beta} \right)+\frac{1}{y}\left(
1-\sqrt{1+2xy\frac{v}{\beta}\left( 1-\frac{v}{\beta} \right)} \right)
\right)}}.
\end{equation}
The condition $|\cos\theta_e| \leq 1$ provides the upper cutoff for
the spectrum.

\begin{figure}
 \begin{center}
  \begin{minipage}{0.85\linewidth}
  \includegraphics[width=1\linewidth]{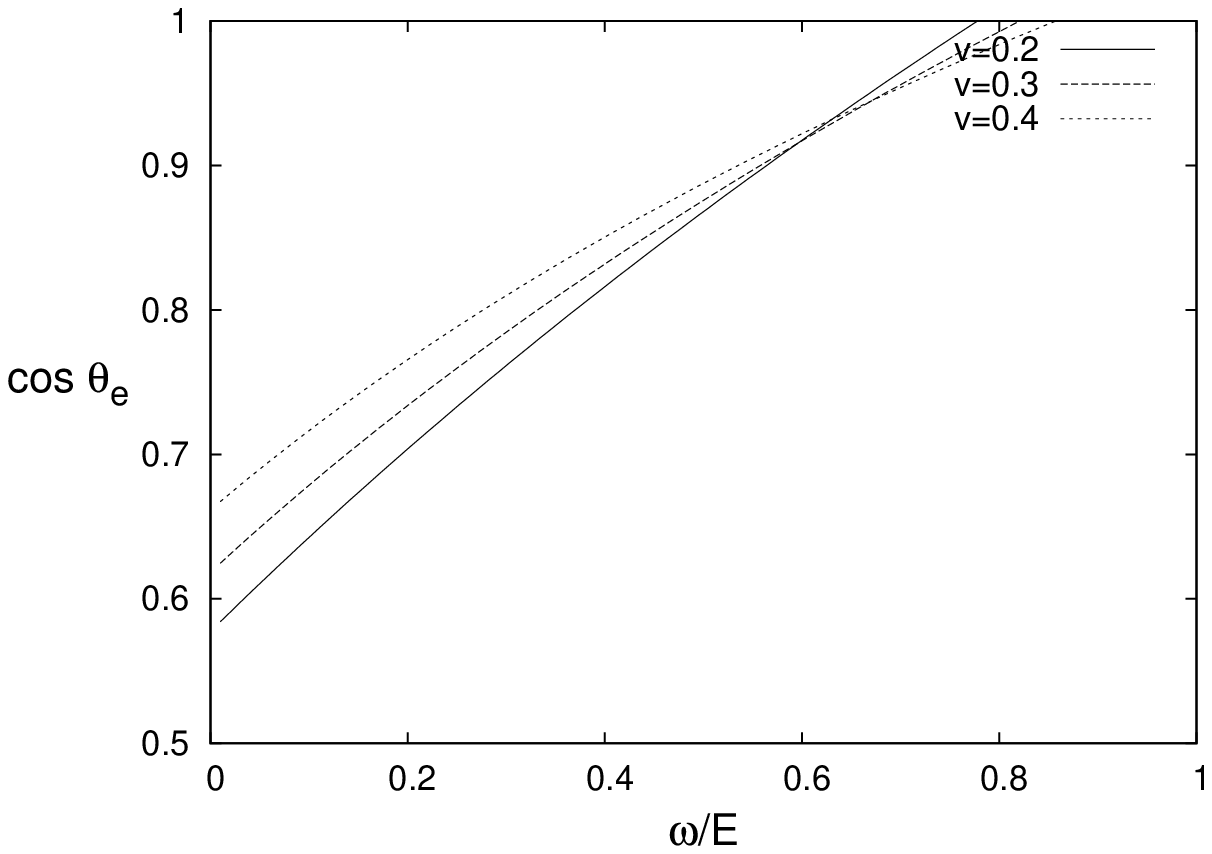}\\
  \caption{Dependence of Cherenkov angle on the energy of emitted photon. $n=2.0$, $\beta=1.0$. Solid line $v=0.2$, dashed line
  $v=0.3$, dotted line $v=0.4$.} \label{ris:angle_1}
  \end{minipage}
 \end{center}
\end{figure}

\begin{figure}
 \begin{center}
  \begin{minipage}{0.85\linewidth}
  \includegraphics[width=1\linewidth]{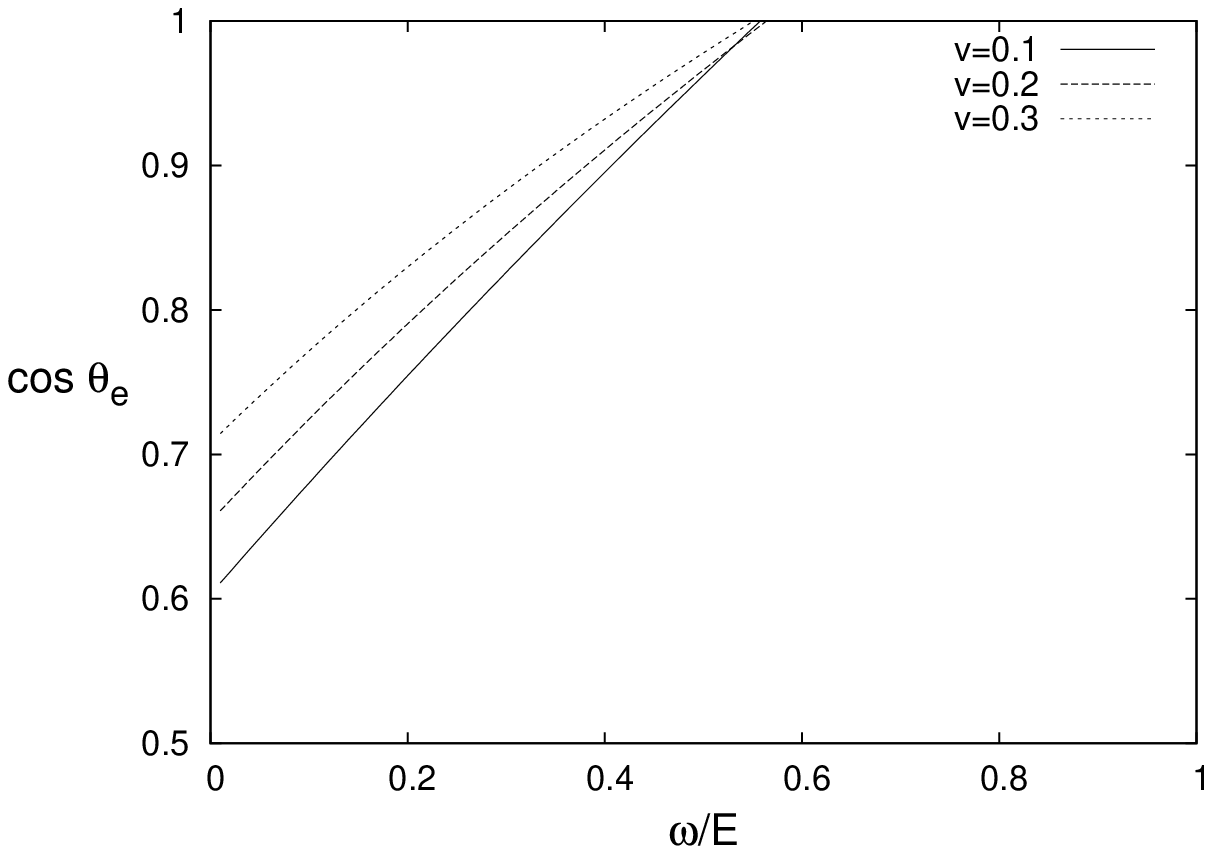}\\
  \caption{Dependence of Cherenkov angle on the energy of emitted photon. $n=2.0$, $\beta=0.9$. Solid line $v=0.1$, dashed line $v=0.2$,
  dotted line $v=0.3$.} \label{ris:angle_2}
  \end{minipage}
 \end{center}
\end{figure}

Let us start with analysis of the dependence of Cherenkov angle on
the energy of the emitted photon. This dependence
(\ref{Cherenkov_angle_1}) is illustrated on the figures
\ref{ris:angle_1} and \ref{ris:angle_2}.

\begin{figure}
 \begin{center}
  \begin{minipage}{0.85\linewidth}
  \includegraphics[width=1\linewidth]{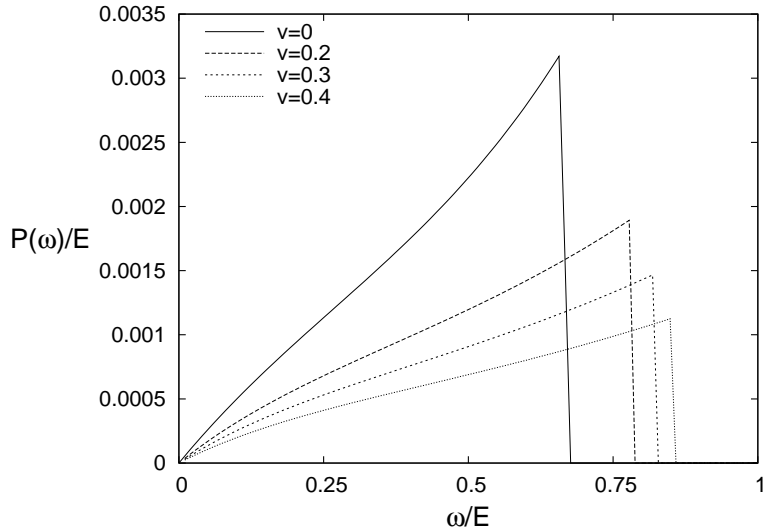}\\
  \caption{Spectrum of Cherenkov radiation in moving medium. $n=2.0$, $\beta=1.0$. Solid line $v=0$, thick dashed line $v=0.2$,
  thin dashed line $v=0.3$, dotted line $v=0.4$.} \label{ris:spectrum1}
  \end{minipage}
 \end{center}
\end{figure}

\begin{figure}
 \begin{center}
  \begin{minipage}{0.85\linewidth}
  \includegraphics[width=1\linewidth]{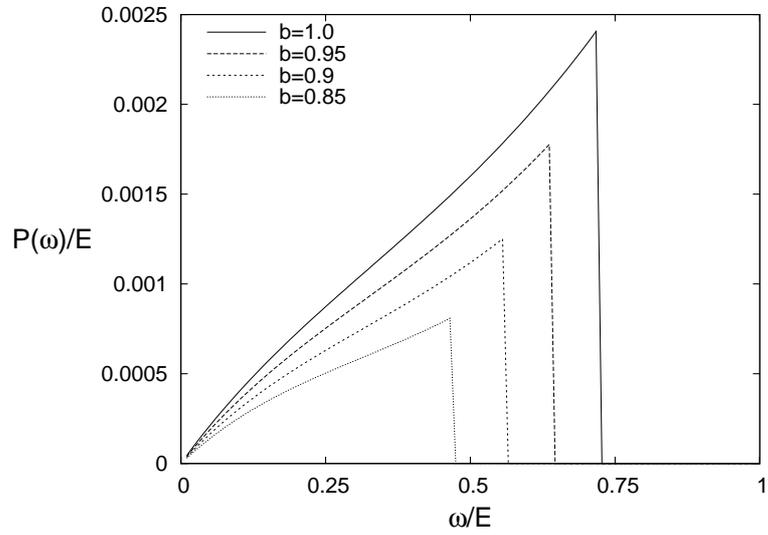}\\
  \caption{Spectrum of Cherenkov radiation in moving medium. $n=2.0$, $v=0.1$. Solid line $\beta=1.0$, thick dashed line $\beta=0.95$,
  thin dashed line $\beta=0.9$, dotted line $v=0.85$.} \label{ris:spectrum2}
  \end{minipage}
 \end{center}
\end{figure}

Then we examine only the spectra for positive $v$ and pay special
attention to the ultra relativistic case ($\beta=1$). These spectra
are pictured on the figures \ref{ris:spectrum1} and
\ref{ris:spectrum2}. One can see the hardening of the spectrum for
$\beta=1$ with the increment of medium speed $v$. This hardening is
explained below.

\begin{figure}
 \begin{center}
  \begin{minipage}{0.85\linewidth}
  \includegraphics[width=1\linewidth]{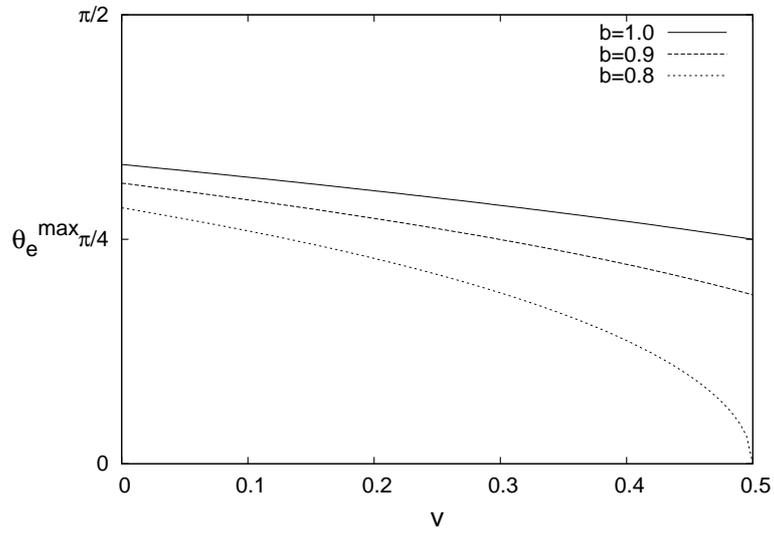}\\
  \caption{Opening angle of the Cherenkov cone. $n=2.0$. Solid line $\beta=1.0$, dashed line $\beta=0.9$,
  dotted line $\beta=0.8$.} \label{ris:opening_angle}
  \end{minipage}
 \end{center}
\end{figure}

There is an important thing to notice about the relationships on the
figures \ref{ris:angle_1} and \ref{ris:angle_2}. All momenta of
radiated Cherenkov photons lie inside the cone with the opening angle
$2\theta_e^{\mathrm{max}}$, corresponding to $x=0$ (Cherenkov photon
with zero energy)

\begin{equation} \label{opening_angle}
\cos\theta_e^{\mathrm{max}}=\frac{1}{\beta\sqrt{1+\frac{n^2-1}{1-v^2}\left(1-\frac{v}{\beta}
\right)^2}}.
\end{equation}
The dependence of this angle on the speed of the medium is
illustrated on the figure \ref{ris:opening_angle}. It can be easily
verified analytically that the function (\ref{opening_angle}) is an
increasing function of $v$. Therefore the opening angle of the
Cherenkov cone decreases as the speed of medium increases.

\begin{figure}
 \begin{center}
  \begin{minipage}{0.85\linewidth}
  \includegraphics[width=1\linewidth]{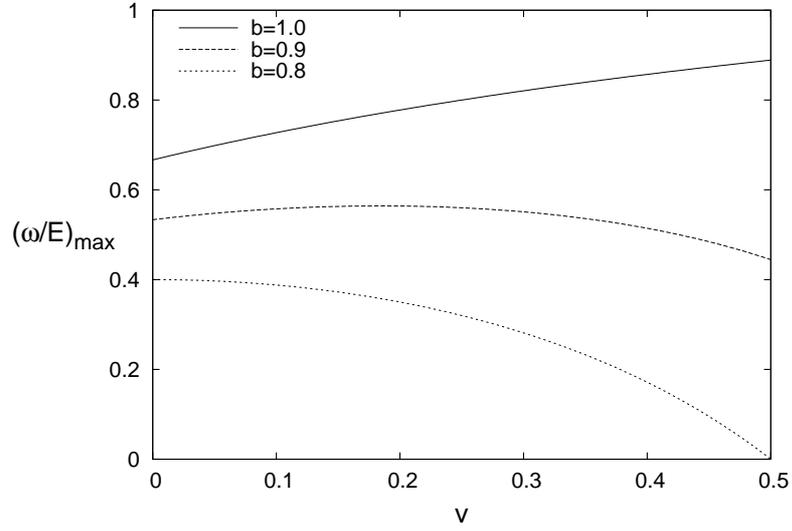}\\
  \caption{The upper bound of the spectrum. $n=2.0$. Solid line $\beta=1.0$, dashed line $\beta=0.9$,
  dotted line $\beta=0.8$.} \label{ris:max}
  \end{minipage}
 \end{center}
\end{figure}

One can see from the figures \ref{ris:angle_1} and \ref{ris:angle_2}
that with increasing photon energy the angle between the photon
momentum and the axis of the cone becomes smaller. Thus the photon
with the maximal energy is radiated with the momentum aligned at the
direction of motion of the fermion. Therefore this energy is the
upper bound of the Cherenkov spectrum. This statement is easily
derived analytically because the function $\cos\theta_e(x)$ is
monotonous. Using the energy-momentum conservation law, we find this
energy (in units of the fermion energy)
\begin{equation}\label{}
x_{max}=\left.\frac{2V_{ph}(\theta)(\beta\cos{\theta}-V_{ph}(\theta))}{1-V_{ph}^2(\theta)}
\right|_{\theta=0}=\frac{2\left(1-\beta\frac{n+v}{1+nv}
\right)}{1-\left(\frac{n+v}{1+nv} \right)^2}.
\end{equation}
This function is a decreasing function of $\beta$ for all $v$,
$|v|<\frac{1}{n}$. We see from the figure \ref{ris:spectrum2} that
the upper bound of the spectrum decreases as the velocity of the
fermion increases. The dependence of the upper bound of the spectrum
on the velocity of the medium is illustrated on the figure
\ref{ris:max}.

>From the figure \ref{ris:max} one can see that in the
ultra-relativistic case ($\beta=1$) the relationship is monotonous,
i.e. the maximal energy of Cherenkov photon increases with growing
$v$. For $\beta=1$ one has
\begin{equation*}
x_{max}=2n-\frac{2(n-1)}{(n+1)(1+v)}.
\end{equation*}
This function is an increasing function of $v$.

\begin{figure}
 \begin{center}
  \begin{minipage}{0.85\linewidth}
  \includegraphics[width=1\linewidth]{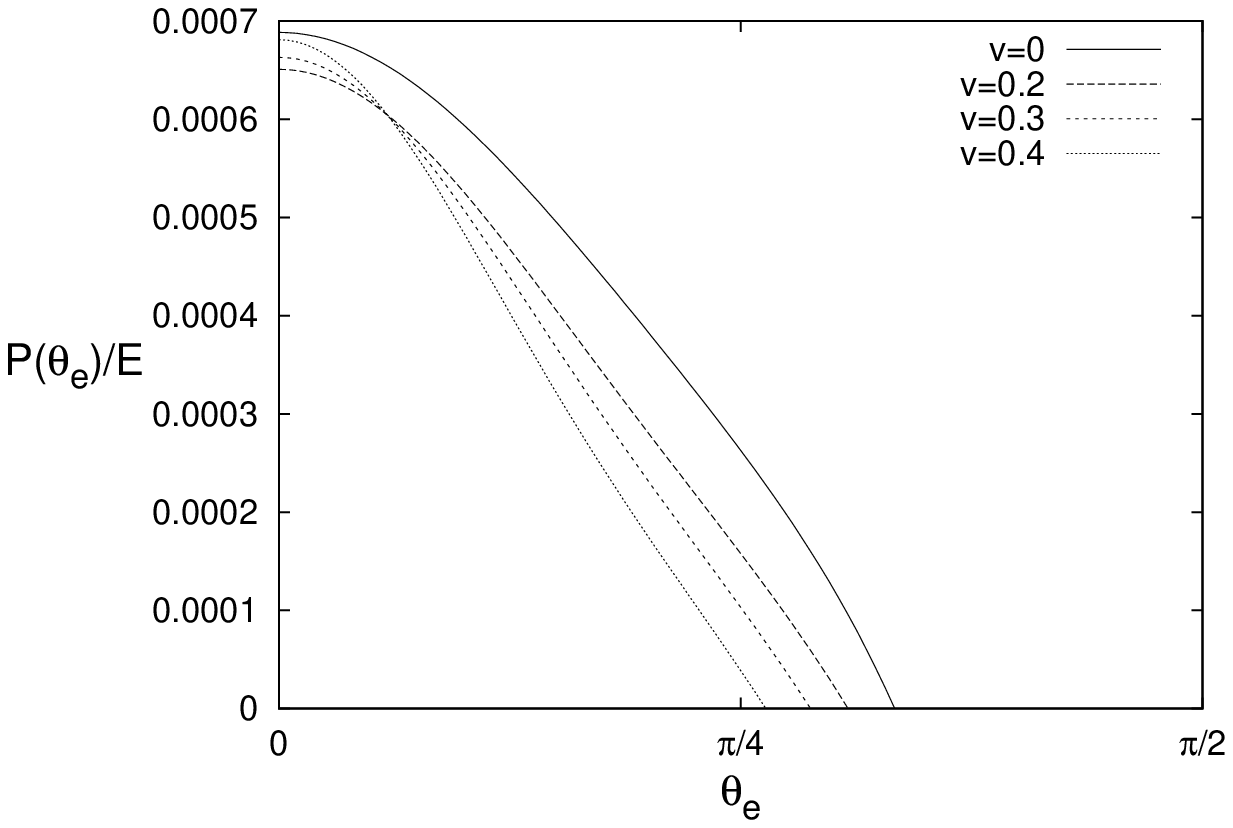}\\
  \caption{Angle distribution of Cherenkov Radiation. $n=2.0$, $\beta=1.0$. Solid line $v=0$, thick dashed line $v=0.2$,
  thin dashed line $v=0.3$, dotted line $v=0.4$.} \label{ris:angle_spectrum_1}
  \end{minipage}
 \end{center}
\end{figure}

\begin{figure}
 \begin{center}
  \begin{minipage}{0.85\linewidth}
  \includegraphics[width=1\linewidth]{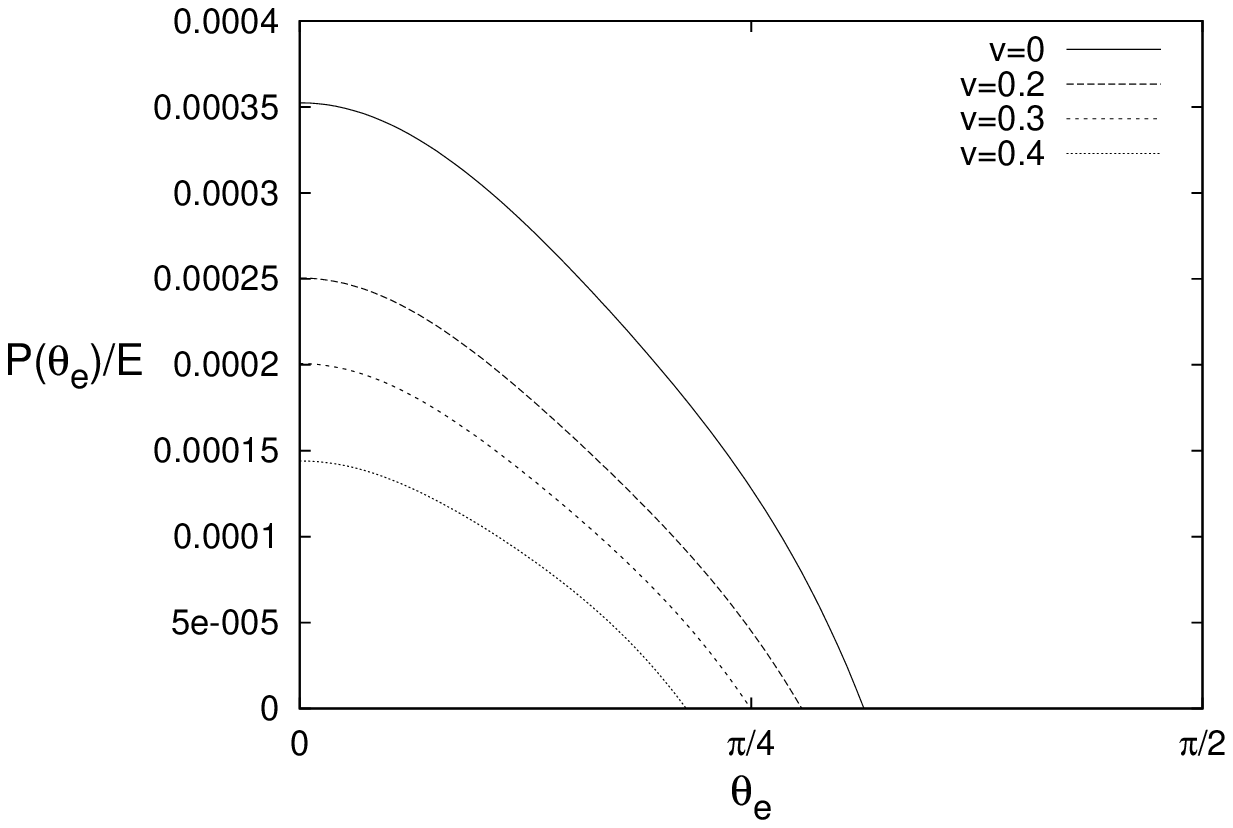}\\
  \caption{Angle distribution of Cherenkov Radiation. $n=2.0$, $\beta=0.9$. Solid line $v=0$, thick dashed line $v=0.2$,
  thin dashed line $v=0.3$, dotted line $v=0.4$.} \label{ris:angle_spectrum_2}
  \end{minipage}
 \end{center}
\end{figure}

Now we can analyze the geometry of the radiation spectrum. Let
$P(\theta_e)$ be energy per unit time radiated in the solid angle
$d\Omega_e=\sin\theta_e d\theta_e d\varphi$, where $\theta_e$ is
Cherenkov angle and $\varphi$ is the polar angle relative to the axis
of fermion's motion. Then it is related with $P(\omega)$ in the
following way
\begin{gather}
P(\theta_e)=\frac{P(\omega(\theta_e))}{2\pi}\frac{\partial
\omega(\theta_e)}{\partial\cos\theta_e}, \\
\omega(\cos\theta_e)=\frac{2V_{ph}(\theta)(\beta\cos{\theta_e-V_{ph}(\theta_e)})}{1-V_{ph}^2(\theta_e)}.
\end{gather}
This spectra are pictured on the figures (\ref{ris:angle_spectrum_1})
and (\ref{ris:angle_spectrum_2}). At first, we notice that the
opening angle of Cherenkov cone decreases as the speed of the medium
increases, which confirms our conclusion. We see that the radiation
is most intensive along the fermion's propagation axis.

Then naturally arises a question why the presented picture of
Cherenkov radiation differs from standard two-bumped structure.
Cherenkov observed a two-bumped structure, but we don't see this
structure in the presented picture of Cherenkov radiation. The origin
of this contradiction is our model of medium. In section 1 we assumed
that the medium doesn't have frequency dispersion and the dielectric
permittivity remains constant for all photon energies $\omega$. But
in real situation this is not true, because the condition
$n(\omega)>1$ can be true only for restricted range of frequencies.
Thus possible explanation of this contradiction is that the energy of
radiating particles is much greater than the energies from this
range. This means that the particle radiates only the photon whose
energy is much less than the energy of the particle. Thus according
to the previous reasonings the momenta of these Cherenkov photons are
close to the surface of the Cherenkov cone. So we obtain the usual
picture of Cherenkov radiation: Cherenkov photons are radiated close
to the surface of Cherenkov cone.

One can examine a lot of interesting cases in moving media. For
example, the situation when speed of medium is greater than phase
speed of light in this medium at rest. It is possible to study this
phenomenon using various relations between velocity of medium and
velocity of a particle. In this article the simplest case was
examined.

\section{Conclusions.}

In this article was conducted the calculation of Cherenkov radiation
spectrum in moving medium with usage of quantum field theory methods.
There were also examined the dependence of this spectrum on the speed
of the medium and the spatial distribution of the radiation. Spectrum
of Cherenkov radiation in moving media differs from the spectrum when
medium is at rest.

At first, the dispersion law of the photon is modified in moving
medium. It reveals different behavior when the speed of medium is
greater or lower than phase speed of light in medium at rest. When
the speed of medium is greater than phase speed of light in medium at
rest there is a region of directions where phase speed in negative.

Let us now summarize the obtained properties of the spectrum. In the
case when the particle and the medium move in the same directions the
spectrum for moving medium lies under the spectrum for medium at
rest. Greater the speed of the medium, lower it lies. But the cut of
the spectrum increases as the speed of the medium increases, and
decreases as the speed of the fermion increases.

The geometrical structure of the Cherenkov radiation of moving medium
has a particular property. All Cherenkov photons are radiated in the
cone with the opening angle $2\theta_e^{\mathrm{max}}$ (where
$\theta_e^{\mathrm{max}}$ is given by the formula
(\ref{opening_angle})), whose axis coincides with the vector of
fermion's momentum. This angle decreases as the speed of the medium
increases. Then the photons with the greater energy have momenta that
are closer to the axis of the cone.

\section*{Acknowledgments}

I am grateful to A.V. Leonidov for suggesting the problem and useful
discussions and I.M. Dremin for useful discussions.

\section*{Appendix}

\appendix

\section{Pole structure}

Let us turn to the pole structure of the expression
(\ref{Cherenkov_moving_spectrum_1}). We write down the denominator
$$(q^2+(n^2-1)(q\widetilde{v})^2+i\varepsilon)((p-q)^2-m^2+i\varepsilon).$$
Let's analyze the expression in the first brackets
\begin{gather*}
(p-q)^2-m^2+i\varepsilon=p^2-2pq+q^2-m^2+i\varepsilon=q^2-2pq+i\varepsilon, \\
(q^0)^2-2Eq^0-|\textbf{q}|^2+2\beta E|\textbf{q}|\cos\theta=0.
\end{gather*}
This equation has two solutions. Taking into account, that because of
$\delta$-function $|\textbf{q}|=\frac{\omega}{V_{ph}}$, we have
\begin{equation}
q^{0}_{1\pm}=E\left\lbrace 1\pm\sqrt{1+\frac{\omega^2}{E^2
V_{ph}^2}-2\beta\frac{\omega}{E V_{ph}}\cos\theta} \right\rbrace\mp
i\varepsilon. \label{pole_1}
\end{equation}

Now we handle the other two poles. We know the solutions of this
expression, so we just write down two residuary poles

\begin{align}
&q^{0}_{2+}=\omega-i\varepsilon \label{pole_2+} \\
&q^{0}_{2-}=\frac{\frac{n^2-1}{1-v^2}v\cos\theta-\left(
1+\frac{n^2-1}{1-v^2}(1-v^2\cos^2 \theta)
\right)^{\frac{1}{2}}}{\frac{n^2-1}{1-v^2}v\cos\theta+\left(
1+\frac{n^2-1}{1-v^2}(1-v^2\cos^2 \theta)
\right)^{\frac{1}{2}}}\omega+i\varepsilon=B(\theta)\omega+i\varepsilon
\label{pole_2-}
\end{align}

Then we must prove two lemmas regarding this poles. But at first we
introduce the designation
\begin{equation}
A(\theta)=\sqrt{1+\frac{\omega^2}{E^2 V_{ph}^2}-2\beta\frac{\omega}{E
V_{ph}}\cos\theta}. \label{A(theta)}
\end{equation}

\textbf{Lemma 1}. It is right for all $\theta$ that
\begin{equation}
A(\theta)\geqslant\sqrt{1-\beta^2}>0.
\end{equation}
\textbf{Proof}.
\begin{gather*}
A(\theta)=\sqrt{1-\beta^2 \cos^2 \theta +\left( \frac{\omega}{E V_{ph}}-\beta\cos\theta \right)^2}\geqslant \\
\geqslant\sqrt{1-\beta^2 \cos^2 \theta}\geqslant\sqrt{1-\beta^2}>0.
\end{gather*}

\textbf{Lemma 2}. For all $\theta$ it is right that
\begin{equation}
   E(1+A(\theta))>\omega.
\end{equation}
\textbf{Proof}. Let's assume the contrary. This means that there
exists $\theta_0$ that
\begin{gather*}
\omega \geqslant E(1+A(\theta_0)) \\
\frac{\omega}{E}\geqslant 1+A(\theta_0)
\end{gather*}
Introducing the designation $x=\frac{\omega}{E}$, we have
\begin{gather*}
x-1\geqslant A(\theta_0)>0, \\
x^2-2x+1 \geqslant 1+\frac{x^2}{V_{ph}^2}-2\beta\frac{x}{V_{ph}}\cos\theta_0, \\
\cos\theta_0 \geqslant\frac{V_{ph}}{\beta} \left( 1+\left( \frac{1}{V_{ph}^2}-1 \right)\frac{x}{2} \right)>\frac{V_{ph}}{\beta}\left( 1+\left( \frac{1}{V_{ph}^2}-1 \right)\frac{1}{2} \right)= \\
=\frac{V_{ph}^2+1}{2\beta V_{ph}}>\frac{1}{\beta}>1.
\end{gather*}
It means that our assumption is not correct and the initial statement
is correct
$$E(1+A(\theta))>\omega.$$

>From the proved lemmas we can make a conclusion that the poles in the
lower half plane never coincide.

\section{Evaluation of the integral}

Let's examine the expressions $D_{1\varepsilon}$ and
$D_{2\varepsilon}$ more carefully. We start from $D_{1\varepsilon}$
\begin{gather*}
D_{1\varepsilon}=(q^0_{1+}-q^0_{1-})(q^0_{1+}-q^0_{2+})(q^0_{1+}-q^0_{2-})= \\
=2(E A(\theta)-i\varepsilon)(E(1+A(\theta))-B(\theta)\omega-2i\varepsilon)(E(1+A(\theta))-\omega) \\
\lim_{\varepsilon\to 0}D_{1\varepsilon}=2E A(\theta)
(E(1+A(\theta))-B(\theta)\omega)(E(1+A(\theta))-\omega).
\end{gather*}
According to \textbf{lemma 1} function $A(\theta)$ is positive for
all $\theta$, and according to \textbf{lemma 2}
$E(1+A(\theta))>\omega$, then these expressions equal 0 nowhere.
Let's write down the function $B(\theta)$

$$B(\theta)=\frac{\frac{n^2-1}{1-v^2}v\cos\theta-\left( 1+\frac{n^2-1}{1-v^2}(1-v^2\cos^2 \theta) \right)^{\frac{1}{2}}}{\frac{n^2-1}{1-v^2}v\cos\theta+\left( 1+\frac{n^2-1}{1-v^2}(1-v^2\cos^2 \theta) \right)^{\frac{1}{2}}}.$$

We will prove that for $|v|<\frac{1}{n}$ the numerator of this
expression is negative.
\begin{gather*}
\frac{n^2-1}{1-v^2}v\cos\theta-\left( 1+\frac{n^2-1}{1-v^2}(1-v^2\cos^2 \theta) \right)^{\frac{1}{2}}<0, \\
\frac{n^2-1}{1-v^2}v\cos\theta<\left( 1+\frac{n^2-1}{1-v^2}(1-v^2\cos^2 \theta) \right)^{\frac{1}{2}}, \\
\frac{(n^2-1)^2}{(1-v^2)^2}v^2 \cos^2 \theta<1+\frac{n^2-1}{1-v^2}(1-v^2\cos^2 \theta), \\
\frac{n^2-1}{1-v^2}v^2 \cos^2 \theta<1
\end{gather*}
It is easy to prove that this inequality is valid for all $\theta$ if
$|v|<\frac{1}{n}$. But this in its part means that the stating
$D_{1\varepsilon}$ does not equal 0 for any $\theta$. Consequently,
$D_{1\varepsilon}$ does not make any contribution to the expression
with an imaginary part.

\section{Calculation of Cherenkov angle}

Let's turn to $D_{2\varepsilon}$.
\begin{gather*}
D_{2\varepsilon}=(q^0_{2+}-q^0_{1+})(q^0_{2+}-q^0_{1-})(q^0_{2+}-q^0_{2-})= \\
=((\omega-E)^2-E^2 A^2 (\theta)+i\varepsilon)(\omega(1-B(\theta))-2i\varepsilon) \\
\lim_{\varepsilon\to 0}D_{2\varepsilon}=((\omega-E)^2-E^2 A^2
(\theta))\omega(1-B(\theta)).
\end{gather*}
Expression in the second brackets doesn't equal 0 because
$B(\theta)<0$ for all $\theta$. And from the stating in the first
brackets we can find the Cherenkov angle. We write down the equation
\begin{equation}
(\omega-E)^2-E^2 A^2 (\theta)=0
\end{equation}
Recalling the equality (\ref{A(theta)}), we obtain the equation
\begin{equation}
\left( \frac{\omega}{E}-1 \right)^2-\left( 1+\frac{\omega^2}{E^2
V_{ph}^2}-2\beta\frac{\omega}{E V_{ph}}\cos\theta \right)=0.
\label{eq1}
\end{equation}
We rewrite the equation (\ref{eq1})
\begin{equation}
(x-1)^2-(1+\frac{x^2}{V_{ph}^2}-2\beta\frac{x}{V_{ph}}\cos\theta)=0.
\label{eq2}
\end{equation}
We will solve this equation in the following way. At first we
transform the equation (\ref{eq2})
\begin{equation}
\cos\theta=\frac{1}{\beta V_{ph}} \left(
V_{ph}^2+(1-V_{ph}^2)\frac{x}{2} \right).
\end{equation}
Thus, we get the system of equations, taking (\ref{Dispersion_law})
into account
$$
\left\{
\begin{array}{lcl}
V_{ph}=\frac{\frac{n^2-1}{1-v^2}v\cos\theta + \left( 1+\frac{n^2-1}{1-v^2}(1-v^2\cos^2 \theta) \right)^{\frac{1}{2}} }{1+\frac{n^2-1}{1-v^2}}, \\
\cos\theta=\frac{1}{\beta V_{ph}} \left( V_{ph}^2+(1-V_{ph}^2)\frac{x}{2} \right). \\
\end{array}
\right.
$$
Now the system takes the form
$$
\left\{
\begin{array}{lcl}
V_{ph}\left( 1+y \right) = yv\cos\theta + \left( 1+y-yv^2\cos^2 \theta \right)^{\frac{1}{2}}, \\
\cos\theta=\frac{1}{\beta V_{ph}} \left( V_{ph}^2+(1-V_{ph}^2)\frac{x}{2} \right). \\
\end{array}
\right.
$$
Then we substitute the second equation into the first one
\begin{multline}
V_{ph}(1+y)-y\frac{v}{\beta V_{ph}}\left( V_{ph}^2 \left( 1-\frac{x}{2} \right)+\frac{x}{2} \right)= \\
=\left( 1+y-y\frac{v^2}{\beta^2 V_{ph}^2}\left( V_{ph}^2 \left(
1-\frac{x}{2} \right)+\frac{x}{2} \right)^2 \right)^{\frac{1}{2}}
\end{multline}
This equation is equivalent to the system of equations
$$
\left\{
\begin{array}{lcl}
V_{ph}^4 \left( 1+y\left( 1-\frac{v}{\beta}\left( 1-\frac{x}{2} \right) \right)^2 \right)-V_{ph}^2 \left( 1+xy\frac{v}{\beta}\left( 1-\frac{v}{\beta}\left( 1-\frac{x}{2} \right) \right) \right)+\frac{1}{4}x^2 y \frac{v^2}{\beta^2}=0, \\
V_{ph}^2\geqslant\frac{xy\frac{v}{\beta}}{2\left( 1+y\left( 1-\frac{v}{\beta}\left( 1-\frac{x}{2} \right) \right) \right)}. \\
\end{array}
\right.
$$
There we must note that following reasonings concern the case $v>0$
(the particle and medium move in the same direction). We solve the
first equation relative to the variable $V_{ph}^2$. Without taking
the inequality for $V_{ph}^2$ into account we obtain two solutions
\begin{equation} \label{V_ph^2}
V_{ph}^2=\frac{x^2 y \frac{v^2}{\beta^2}}{2\left(
1+xy\frac{v}{\beta}\left( 1-\frac{v}{\beta}\left( 1-\frac{x}{2}
\right) \right)\pm\sqrt{1+2xy\frac{v}{\beta}\left( 1-\frac{v}{\beta}
\right)} \right)}
\end{equation}
Then we substitute these solutions in the inequality
\begin{gather*}
\frac{x^2 y \frac{v^2}{\beta^2}}{2\left( 1+xy\frac{v}{\beta}\left( 1-\frac{v}{\beta}\left( 1-\frac{x}{2} \right) \right)\pm\sqrt{1+2xy\frac{v}{\beta}\left( 1-\frac{v}{\beta} \right)} \right)}\geqslant \\
\geqslant\frac{xy\frac{v}{\beta}}{2\left( 1+y\left( 1-\frac{v}{\beta}\left( 1-\frac{x}{2} \right) \right) \right)}, \\
1-\frac{v}{\beta}x \pm \sqrt{1+2xy\frac{v}{\beta}\left(
1-\frac{v}{\beta} \right)} \leqslant 0.
\end{gather*}

If one brings the considered inequality with the sign "$-$" it is
deliberately valid for all $\theta$. If one takes it with the sign
"$+$" this becomes essential to do some work to prove that it is not
valid for all $\theta$. Let's write down the second equation of the
system
$$
x=\frac{2V_{ph}}{1-V_{ph}^2}(\beta\cos\theta-V_{ph}),
$$
from which it is obvious that cherenkov photon is emitted if the
projection of particle's speed on this direction is greater than
phase speed of light in this direction. Let's assume that $x
\geqslant 1$, then
$$
\frac{2V_{ph}}{1-V_{ph}^2}(\beta\cos\theta-V_{ph}) \geqslant 1,
$$
which leads to
$$
(V_{ph}-\beta\cos\theta)^2+1-\beta^2 \cos^2 \theta \leqslant 0,
$$
which is not right for all $\theta$. Previous argumentation proves
obvious physical statement that a particle cannot emit a photon with
an energy greater than its own. We write the inequality allowing for
$x<1$
$$
1-\frac{v}{\beta}x+\sqrt{1+2xy\frac{v}{\beta}\left( 1-\frac{v}{\beta}
\right)} \leqslant 0.
$$
It is not right for all $\theta$ because its left part is less than
1. Thus, only one solution remains
\begin{equation}
V_{ph}=\frac{x \frac{v}{\beta} \sqrt{y}}{\sqrt{2\left(
1+xy\frac{v}{\beta}\left( 1-\frac{v}{\beta}\left( 1-\frac{x}{2}
\right) \right)-\sqrt{1+2xy\frac{v}{\beta}\left( 1-\frac{v}{\beta}
\right)} \right)}}
\end{equation}
After rather boring calculations we get the expression for
$\cos\theta$ in the case $v>0$
\begin{equation} \label{Cherenkov_angle}
\cos\theta_e=\frac{1+\frac{\beta}{xyv}\left(
1-\sqrt{1+2xy\frac{v}{\beta}\left( 1-\frac{v}{\beta} \right)}
\right)}{\sqrt{v^2+\frac{2\beta^2}{x^2}\left( \frac{xv}{\beta}\left(
1-\frac{v}{\beta} \right)+\frac{1}{y}\left(
1-\sqrt{1+2xy\frac{v}{\beta}\left( 1-\frac{v}{\beta} \right)} \right)
\right)}}.
\end{equation}

\end{document}